\documentclass[12pt,prd,showpacs,tightenlines,nofootinbib]{revtex4}
\usepackage{bm}
\usepackage{graphics}
\usepackage{rotating}
\usepackage{epsfig}
\begin{document}
\title{Strong decays of vector mesons 
to pseudoscalar mesons 
in the relativistic quark model}
\author{D. Ebert$^{1}$, R. N. Faustov$^{2}$  and V. O. Galkin$^{2}$}
\affiliation{
$^1$ Institut f\"ur Physik, Humboldt--Universit\"at zu Berlin,
Newtonstrasse 15, D-12489  Berlin, Germany\\
$^2$ Dorodnicyn Computing Centre, Russian Academy of Sciences,
  Vavilov Street 40, 119333 Moscow, Russia}
\begin{abstract}
Strong decays of vector ($^3S_1$) mesons to the
pair of pseudoscalar ($^1S_0$) mesons are considered in the
framework of the microscopic decay mechanism and the relativistic quark model
based on the quasipotential approach. The quark-antiquark potential,
which was previously used for the successful description of meson
spectroscopy and electroweak decays, is employed as the source of the
$q\bar q$ pair creation. The relativistic structure of the decay matrix
element, relativistic contributions and boosts of the meson wave
functions are comprehensively taken into account. The calculated
rates of strong decays of light, heavy-light mesons and heavy quarkonia agree well
with available experimental data.  
\end{abstract}


\maketitle

\section{Introduction}
At present a large set of experimental data is available on light and
heavy mesons which is constantly extending \cite{pdg}. Recently, the most dramatic
progress has been achieved in 
the heavy meson sector. As a result, in the last
years many new charmonium states and new bottomonium states have been
discovered \cite{qwg}. The number of the open-flavour (charmed $D$ and bottom
$B$) meson states is also constantly increasing. 
Some of the new states  are
the long-awaited ones, expected within the constituent quark model many 
years ago, while some others, with masses higher than the
thresholds of the open charm and bottom production, have narrow widths and
unexpected decay properties \cite{pdg,qwg}. Similar exotic states are
known in the light meson sector. There are theoretical indications \cite{qwg}
that some of these states could be the first manifestation of the
existence of 
exotic hadrons (tetraquarks, molecules, hybrids etc.), which are
predicted to exist within quantum chromodynamics (QCD).
In order to explore such entities, a comprehensive understanding of the
meson spectroscopy and decays up to
rather high orbital and radial excitations is required. 

Thus, one of the important issues is the study of  
strong meson decays. Such decays are the main channels for
mesons with masses above open flavour production thresholds. 
They are investigated already for many 
years. Nevertheless, strong decays still constitute a rather poorly
understood area of hadronic physics in view of their complex
nonperturbative dynamics, which has not yet been deduced directly from QCD.          

Several phenomenological models of open-flavour strong decays have been
proposed and described in the literature.
Some of them are based on effective chiral meson Lagrangians 
derived from the Nambu-Jona-Lasinio quark model (see e.g. \cite{erv}
and references therein),
which express decay amplitudes through quark loop integrals with quark
propagators between initial and final meson vertices. 
However, such considerations of strong decays do not account
for quark confinement and momentum-dependence of vertices. 
In particular,
  vertices are considered to be point-like and, as a result, 
quark loop integrals diverge. Thus, the 
introduction of some phenomenological
cutoff parameter is necessary.   

The other group of approaches are based on different types of  quark
pair creation models. They differ in the production mechanism of a
light
quark pair from the QCD vacuum. The phenomenological $^3S_1$ model
\cite{3s1} considers the corresponding quark-antiquark pair to be
produced in the vector state. However, this appears to disagree with
experiment and is thus ruled out \cite{gs}. 

The most popular approach to strong meson decays is based on  
the phenomenological $^3P_0$ model (see \cite{3p0,gs,qwg} and
references therein). This model 
assumes that the quark-antiquark pair is created with the vacuum
quantum numbers, $J^{PC}=0^{++}$. It gives for most decays 
results in fairly good
agreement with experimental data. The important features of the model
are its simplicity and necessity to introduce only one additional
parameter, the strength of the decay interaction, in order to describe various
strong decays. This parameter is considered as a free constant and
is fitted to the data. It is generally believed that the pair
production strength parameter is roughly flavour independent, but
recent studies, involving a
global fit of the experimental data,
indicate that it can be scale dependent \cite{sef}. However, 
this model
does not clarify the fundamental mechanism of pair creation. It has
an
explicitly nonrelativistic character with meson wave functions modeled
by simple Gaussian functions. All this makes it very difficult
to improve the model.   
 
Another approach, closely related to  the $^3P_0$  model, 
is the flux-tube breaking model \cite{ki}. It also assumes that a quark-antiquark pair is created with the
vacuum quantum numbers, but it additionally includes
the overlaps of the flux-tube of the initial meson with those of the
two final mesons. Therefore, the resulting calculations are more
complicated, but lead to predictions close to the results of  the $^3P_0$  model.
  
In the microscopic decay model \cite{abs}, which is more closely
related to QCD, the pair creation
originates from the current-current interactions due to the potential
binding quarks in mesons, 
which is usually assumed to be the sum of the
scalar confining interaction and the one gluon exchange. It
generalizes the
above approaches, which can be obtained as its special
limiting cases. Thus, 
the $^3P_0$ model results are reproduced with the
constant scalar interaction. In contrast to 
the $^3P_0$ model, the strong decay rates are completely determined by the
meson wave functions, quark masses and interaction parameters. At
present, most
calculations are done in the nonrelativistic constituent quark model
with spherical harmonic oscillator wave functions.

In Ref.~\cite{sim} a model of strong decays 
has been proposed, where
the pair creation occurs due to the string breaking. The basic
interaction in this model is the scalar colour-singlet confining
potential acting between the light quark and heavy antiquark. It is flavour independent and  nonlocal for the zero mass
light quark pair, turning to the linearly rising confining potential
for a long breaking string. There is a direct  correspondence between
the string  breaking model and the microscopic decay model with the
scalar potential, but the former  uses a
relativistic formalism for
light quarks with vanishing current masses.    

In this paper we propose a
relativistic approach for
the calculation of 
strong decays of mesons in the framework of the
previously developed relativistic quark model \cite{mass,fg,sbar}
based on the quasipotential approach. For this purpose, 
the microscopic
decay model is extended to include relativistic effects into decay
matrix elements, 
relativistic corrections and boosts of the meson wave functions. The
QCD-motivated quark-antiquark interaction potential, 
which was previously found to reproduce well mass spectra and
electroweak decays of mesons, is used for the description of the
pair creation mechanism. The resulting
relativistic calculations are rather complicated. 
As a first step, we
consider the simplest case,
 where only  $S$-wave mesons are
involved. This significantly simplifies the angular structure of decay
matrix elements. 
The comparison 
of the obtained results with the available
experimental data for the decays of vector ($^3S_1$) mesons to a
pair of pseudoscalar ($^1S_0$) mesons provides a
test of our approach.

\section{Relativistic quark model}  
\label{rqm}

For the following calculations we use the relativistic quark
model  based on the
quasipotential approach and quantum chromodynamics (QCD).  Mesons are
considered as the bound states of constituent quarks which are described by the
single-time wave functions satisfying the
three-dimensional relativistically invariant Schr\"odinger-like
equation with the QCD-motivated interquark potential \cite{mass}  
\begin{equation}
\label{quas}
{\left(\frac{b^2(M)}{2\mu_{R}}-\frac{{\bf
p}^2}{2\mu_{R}}\right)\Psi_{M}({\bf p})} =\int\frac{d^3 q}{(2\pi)^3}
 V({\bf p,q};M)\Psi_{M}({\bf q}),
\end{equation}
with  the relativistic reduced mass defined by
\begin{equation}
\mu_{R}=\frac{M^4-(m^2_1-m^2_2)^2}{4M^3},
\end{equation}
where $M$ is the meson mass, $m_{1,2}$ are the quark masses,
and ${\bf p}$ is the relative momentum of the constituent quarks.  
In the center of mass system the relative momentum squared on the mass shell 
$b^2(M)$ is expressed through the meson and quark masses:
\begin{equation}
{b^2(M) }
=\frac{[M^2-(m_1+m_2)^2][M^2-(m_1-m_2)^2]}{4M^2}.
\end{equation}
It is assumed that the kernel of this equation -- the interquark quasipotential $V({\bf
  p,q};M)$ -- consists of the perturbative one-gluon
exchange (OGE) and the nonperturbative confining parts \cite{mass}
  \begin{equation}
\label{qpot}
V({\bf p,q};M)=\bar{u}_1(p)\bar{v}_2(-p){\mathcal V}({\bf p}, {\bf
q};M)u_1(q)v_2(-q), 
\end{equation}
with
$${\mathcal V}({\bf p},{\bf q};M)={\mathcal V}({\bf k})=\frac{4}{3}\alpha_sD_{ \mu\nu}({\bf
k})\gamma_1^{\mu}\gamma_2^{\nu}
+V^V_{\rm conf}({\bf k})\Gamma_1^{\mu}({\bf k})
\Gamma_{2;\mu}({\bf k})+V^S_{\rm conf}({\bf k}),$$
where ${\bf k=p-q}$, $\alpha_s$ is the QCD coupling constant, $D_{\mu\nu}$ is the
gluon propagator in the Coulomb gauge, while $\gamma_{\mu}$ and
$u_{1}$, $v_{2}$
are  the Dirac matrices and spinors, respectively. 

The confining part is taken as the mixture of
the Lorentz-scalar and Lorentz-vector linearly
rising interactions which in the nonrelativistic limit reduce to
\begin{equation}
\label{nr}
V_{\rm conf}(r)=V_{\rm conf}^S(r)+V_{\rm conf}^V(r)=Ar+B,
\end{equation}
with
\begin{equation}
\label{vlin}
V_{\rm conf}^V(r)=(1-\varepsilon)(Ar+B),\qquad 
V_{\rm conf}^S(r) =\varepsilon (Ar+B).
\end{equation}
Therefore, in this limit the Cornell-type potential is reproduced
$$V_{\rm NR}(r)=-\frac43\frac{\alpha_s}{r}+Ar+B.$$
The value of the mixing coefficient $\varepsilon=-1$
has been obtained from the consideration of the heavy quark expansion
for the semileptonic $B\to D^{(*)}$ decays
\cite{fg} and charmonium radiative decays \cite{mass}.

For the QCD coupling constant  $\alpha_s\equiv\alpha_s(\mu^2)$
we set the scale $\mu=2m_1 m_2/(m_1+m_2)$ and use the
model with freezing \cite{bvb}
\begin{equation}
  \label{eq:alpha}
  \alpha_s(\mu^2)=\frac{4\pi}{\displaystyle\beta_0
\ln\frac{\mu^2+M_B^2}{\Lambda^2}}, \qquad \beta_0=11-\frac23n_f,
\end{equation}
where the background mass is $M_B=2.24\sqrt{A}=0.95$~GeV, and
$\Lambda=413$~MeV was fixed in our model from fitting light and
heavy-light meson spectra \cite{mass}.

The vector vertex of the confining interaction contains the additional Pauli
term with the nonperturbative anomalous chromomagnetic moment of the quark $\kappa$
\begin{equation}
\label{kappa}
\Gamma_{\mu}({\bf k})=\gamma_{\mu}+
\frac{i\kappa}{2m}\sigma_{\mu\nu}k^{\nu}.
\end{equation}
We fixed the value  $\kappa=-1$  by
analyzing the fine splittings of heavy quarkonia ${}^3P_J$- states \cite{mass} and  the heavy quark expansion for semileptonic
decays of heavy mesons \cite{fg} and baryons \cite{sbar}. It
enables the
vanishing of the spin-dependent chromomagnetic interaction,
proportional to $(1+\kappa)$, in accord with the flux tube model. 
The constituent quark masses $m_b=4.88$ GeV, $m_c=1.55$ GeV, $m_s=0.5$ GeV, $m_{u,d}=0.33$ GeV and
the parameters of the linear potential  $A=0.18$ GeV$^2$ and
$B=-0.30$ GeV  were determined from the previous analysis of
meson spectroscopy  \cite{mass}. Note that we have used a
universal set of 
model parameters for the calculations of the meson, baryon and
tetraquark spectra as well as their weak and radiative decays.  

In the following section we 
apply our relativistic quark model to
the consideration of the strong decays of vector mesons to 
a pair of pseudoscalar mesons.

\section{Relativistic description of strong decays in  
a microscopic decay model}

The current-current interaction in the microscopic decay model
\cite{abs} is described by the following Hamiltonian
\begin{equation}
  \label{eq:hi}
  H_I=\frac12\int\int d^3 x d^3 y J({\bf x})\frac{\lambda^a}2V(|{\bf x}-{\bf y}|)J({\bf y})\frac{\lambda^a}2, 
\end{equation}
where in our model the quark current $J$ is given by
\begin{equation}
  \label{eq:J}
  J\equiv \bar \psi\Gamma \psi=\left\{\begin{array}{ll}\bar \psi \psi&{\rm scalar\ confining\
        interaction},\\ \\ \bar \psi(\gamma^\mu+i\frac{\kappa}{2m}\sigma^{\mu\nu}k_\nu) \psi&{\rm vector\ confining\
        interaction\ with \ the \ Pauli \ term},\\ \\ \bar \psi\gamma^0 \psi & {\rm color\ Coulomb\
        OGE},\\ \\
  (\bar\psi\gamma^i \psi)_T & {\rm transverse\
        OGE},
\end{array}\right.
\end{equation}
and the interaction kernel $V$, according to Ref.~\cite{abs}, is defined as
\begin{equation}
  \label{eq:V}
  V(r)=\left\{\begin{array}{ll}\frac34\varepsilon(Ar+B)&{\rm scalar\ confining\
        interaction},\\ \\ \frac34(1-\varepsilon)(Ar+B)&{\rm vector\ confining\
        interaction},\\ \\ \alpha_s/r & {\rm color\ Coulomb\
        OGE},\\ \\
 -\alpha_s/r & {\rm transverse\
        OGE}.
\end{array}\right.
\end{equation}
Here the confinement kernel is normalized so that one
gets the confining potentials in Eq.~(\ref{nr}) of a color-singlet $q\bar q$
pair.

\begin{figure}

  \includegraphics[width=6cm]{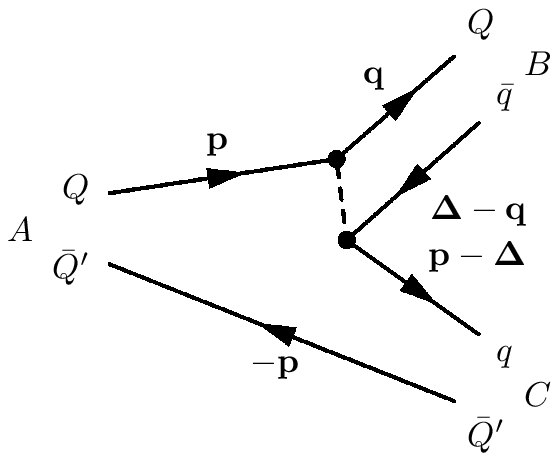} \qquad  \includegraphics[width=6cm]{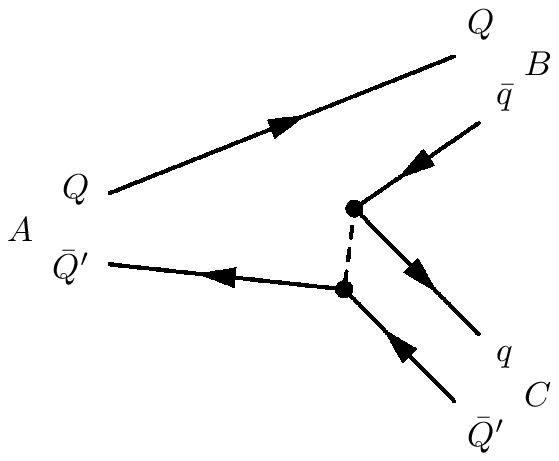}\\
($d1$)\hspace*{6cm} ($d2$)

\caption{\label{fig:1a} Diagrams of the strong decay $A\to BC$.  }
\end{figure}

The strong decay  process $A\to BC$, where mesons have the following
quark content: $A(Q\bar Q')$, $B(Q\bar
q)$, $C(q\bar Q')$, is described by the two
diagrams given
in Fig.~\ref{fig:1a}. The corresponding decay matrix element can be presented as  
\begin{equation}
\label{eq:mel}
\langle BC|H_I|A\rangle=h_{fi}\ \delta({\bf P}_A-{\bf P}_B-{\bf P}_C),
\end{equation}
where ${\bf P}_I$ $(I=A,B,C)$ are three-momenta of mesons and the
$\delta$-function accounts for the momentum conservation. The
 matrix element $h_{fi}$ is the product of the Fermi signature 
phase (due to
 permutation of quark and antiquark operators) $I_{\rm signature}$
and the color $I_{\rm color}$,
flavour 
$I_{\rm flavor}$ and spin-space $I_{\rm spin-space}$ factors 
\begin{equation}
  \label{eq:hh}
  h_{fi}=I_{\rm signature}\ I_{\rm color}\ I_{\rm flavor}\ I_{\rm spin-space}.
\end{equation}
The expressions for the factors $I_{\rm signature}$, $I_{\rm color}$ and  $I_{\rm flavor}$ can be found
in Ref.~\cite{abs}. 

The spin-space factor $I_{\rm spin-space}$ 
can be expressed through
the overlap integral of the meson wave functions. In the rest frame of
the decaying meson $A$ (${\bf P}_A=0$, ${\bf P}_B=-{\bf  P}_C={\bf
  \Delta}$) the contribution of the $(d1)$ diagram in Fig.~\ref{fig:1a}
is given by 
\begin{eqnarray}
  \label{eq:isp}
  I_{\rm spin-space}(d1)&
=&\int\int \frac{d^3p d^3q}{(2\pi)^3}\bar\Psi_{B\,{\bf \Delta}}(2{\bf q}-{\bf
  \Delta})\bar\Psi_{C\,-{\bf \Delta}}(2{\bf p}-{\bf \Delta}) [\bar
  u_{Q}(q)\Gamma u_Q(p)]\cr
&&\times{\cal V}({\bf p}-{\bf q})[\bar
u_q(p-\Delta)\Gamma v_q(-q+\Delta)]\Psi_{A\,{\bf 0}}(2{\bf p}).
\end{eqnarray}
It is important to note that the wave functions entering the decay
matrix element (\ref{eq:isp}) are not in the rest frame. 
In the chosen frame, where the initial vector $A$ meson is at rest
(${\bf P}_{A}=0$), the final $B$ and $C$  mesons
are moving with the recoil momenta ${\bf P}_B=-{\bf
  P}_C={\bf \Delta}$. The wave function
of the moving 
$M(q_1\bar q_2)$  meson $\Psi_{M\,{\bf\Delta}}$ is connected 
with the  wave function in the rest frame 
$\Psi_{M\,{\bf 0}}\equiv \Psi_M$ by the transformation \cite{f}
\begin{equation}
\label{wig}
\Psi_{M\,{\bf\Delta}}({\bf
p})=D_{q_1}^{1/2}(R_{L_{\bf\Delta}}^W)D_{q_2}^{1/2}(R_{L_{
\bf\Delta}}^W)\Psi_{M\,{\bf 0}}({\bf p}),
\end{equation}
where $q_1$, $q_2$ denote $Q(Q')$ or $q$; $R^W$ is the Wigner rotation, $L_{\bf\Delta}$ is the Lorentz boost
from the meson rest frame to a moving one, and   
the rotation matrix $D^{1/2}(R)$ in spinor representation is given by
\begin{equation}\label{d12}
{1 \ \ \,0\choose 0 \ \ \,1}D^{1/2}_{q_i}(R^W_{L_{\bf\Delta}})=
S^{-1}({\bf p}_{q_i})S({\bf\Delta})S({\bf p}), \qquad (i=1,2),
\end{equation}
where
$$
S({\bf p})=\sqrt{\frac{\epsilon(p)+m}{2m}}\left(1+\frac{\bm{\alpha}{\bf p}}
{\epsilon(p)+m}\right)
$$
is the usual Lorentz transformation matrix of the four-spinor and
$\epsilon(p)=\sqrt{m^2+{\bf p}^2}$ is the
quark energy.

Substituting corresponding spinors in Eq.~(\ref{eq:isp}) and taking
into account the wave function transformations (\ref{wig}) and spin
structure of the initial vector ($^3S_1$) and final pseudoscalar
($^1S_0$) mesons, we get  the spin-space factor in the form 
\begin{equation}
  \label{eq:isp1}
  I_{\rm spin-space}(d1)=
\int\int \frac{d^3p d^3q}{(2\pi)^3} \bar\Psi_{B,{\bf 0}}(2{\bf q}-{\bf
  \Delta})\bar\Psi_{C,{\bf 0}}(2{\bf p}-{\bf \Delta}){\cal M}({\bf p},{\bf q}) {\cal V}({\bf p}-{\bf q})
\Psi_{A,{\bf 0}}(2{\bf p}),
\end{equation}
with the functions $ {\cal M}({\bf p},{\bf q})$ given by the following
expressions.

(a) scalar confining interaction $\Gamma=I$
\begin{eqnarray}
  \label{eq:msc}
  {\cal M}_{\rm
    scal}({\bf p},{\bf q})&=&\sqrt{\frac{\epsilon_Q(q)+m_Q}{2\epsilon_Q(q)}}\sqrt{\frac{\epsilon_Q(p)+m_Q}{2\epsilon_Q(p)}}
  \sqrt{\frac{\epsilon_q(p-\Delta)+m_q}{2\epsilon_q(p-\Delta)}} 
\sqrt{\frac{\epsilon_q(q-\Delta)+m_q}{2\epsilon_q(q-\Delta)}}\cr
&&\Biggl[p_+\Biggl\{-\frac1{\epsilon_q(p-\Delta)+m_q}\left(1-
  \frac12\frac{{\bf
      \Delta}^2-(E_B-M_B)(E_C-M_C)}{(E_C+M_C)(\epsilon_{\bar
     Q'}(p-\Delta)+m_{\bar Q'})}\right)\cr
&&+\frac12\frac{{\bf
      \Delta}^2+(E_B+M_B)(E_C-M_C)}{(E_C+M_C)(\epsilon_{\bar
     Q'}(p-\Delta)+m_{\bar Q'})(\epsilon_q(q-\Delta)+m_q)}\Biggr\}\cr
&&+q_+\Biggl\{-\frac1{\epsilon_q(q-\Delta)+m_q}\left(1-
  \frac12\frac{{\bf
      \Delta}^2+(E_B+M_B)(E_C-M_C)}{(E_B+M_B)(\epsilon_{
      Q}(q-\Delta)+m_{Q})}\right)\cr
&&+\frac12\frac{{\bf
      \Delta}^2-(E_B-M_B)(E_C-M_C)}{(E_B+M_B)(\epsilon_{
      Q}(q-\Delta)+m_{Q})(\epsilon_q(p-\Delta)+m_q)}\Biggr\}\Biggr],
\end{eqnarray}

(b) vector confining interaction with the Pauli term
$\Gamma=\gamma^\mu+i\frac{\kappa}{2m}\sigma^{\mu\nu}k_\nu$ for
$\kappa=-1$
\begin{eqnarray}
  \label{eq:mvc}
  {\cal M}_{\rm
    vect}({\bf p},{\bf q})&=&\sqrt{\frac{\epsilon_Q(q)+m_Q}{2\epsilon_Q(q)}}\sqrt{\frac{\epsilon_Q(p)+m_Q}{2\epsilon_Q(p)}}
  \sqrt{\frac{\epsilon_q(p-\Delta)+m_q}{2\epsilon_q(p-\Delta)}} 
\sqrt{\frac{\epsilon_q(q-\Delta)+m_q}{2\epsilon_q(q-\Delta)}}\cr
&&\Biggl[p_+\Biggl\{-\frac1{\epsilon_Q(p-\Delta)+m_Q}\left(1-
 \frac{M_B(E_C-M_C)}{(E_C+M_C)(\epsilon_{\bar
     Q'}(p-\Delta)+m_{\bar Q'})}\right)\Biggr\}\cr
&&+q_+\Biggl\{-\frac1{\epsilon_Q(q-\Delta)+m_Q}\left(1-
  \frac{M_B(E_C-M_C)}{(E_B+M_B)(\epsilon_{
      Q}(q-\Delta)+m_{Q})}\right)\Biggr\}\Biggr],
\end{eqnarray}

(c) color Coulomb OGE $\Gamma=\gamma^0$
\begin{eqnarray}
  \label{eq:mcc}
 {\cal M}_{\rm
    Coul}({\bf p},{\bf q})&=&\sqrt{\frac{\epsilon_Q(q)+m_Q}{2\epsilon_Q(q)}}\sqrt{\frac{\epsilon_Q(p)+m_Q}{2\epsilon_Q(p)}}
  \sqrt{\frac{\epsilon_q(p-\Delta)+m_q}{2\epsilon_q(p-\Delta)}} 
\sqrt{\frac{\epsilon_q(q-\Delta)+m_q}{2\epsilon_q(q-\Delta)}}\cr
&&\Biggl[p_+\Biggl\{\frac1{\epsilon_q(p-\Delta)+m_q}\left(1-
  \frac12\frac{{\bf
      \Delta}^2-(E_B-M_B)(E_C-M_C)}{(E_C+M_C)(\epsilon_{\bar
     Q'}(p-\Delta)+m_{\bar Q'})}\right)\cr
&&-\frac12\frac{{\bf
      \Delta}^2+(E_B+M_B)(E_C-M_C)}{(E_C+M_C)(\epsilon_{\bar
     Q'}(p-\Delta)+m_{\bar Q'})(\epsilon_q(q-\Delta)+m_q)}\Biggr\}\cr
&&+q_+\Biggl\{-\frac1{\epsilon_q(q-\Delta)+m_q}\left(1-
  \frac12\frac{{\bf
      \Delta}^2+(E_B+M_B)(E_C-M_C)}{(E_B+M_B)(\epsilon_{
      Q}(q-\Delta)+m_{Q})}\right)\cr
&&+\frac12\frac{{\bf
      \Delta}^2-(E_B-M_B)(E_C-M_C)}{(E_B+M_B)(\epsilon_{
      Q}(q-\Delta)+m_{Q})(\epsilon_q(p-\Delta)+m_q)}\Biggr\}\Biggr],
\end{eqnarray}

(d) transverse OGE $\Gamma=\gamma^i_T$
\begin{eqnarray}
  \label{eq:mtc}
 {\cal M}_{T}({\bf p},{\bf q})&=&\sqrt{\frac{\epsilon_Q(q)+m_Q}{2\epsilon_Q(q)}}\sqrt{\frac{\epsilon_Q(p)+m_Q}{2\epsilon_Q(p)}}
  \sqrt{\frac{\epsilon_q(p-\Delta)+m_q}{2\epsilon_q(p-\Delta)}} 
\sqrt{\frac{\epsilon_q(q-\Delta)+m_q}{2\epsilon_q(q-\Delta)}}\cr
&&\Biggl[p_+\Biggl\{\frac1{\epsilon_Q(p-\Delta)+m_Q}\left(1+
  \frac12\frac{{\bf
      \Delta}^2+(E_B-M_B)(E_C-M_C)}{(E_C+M_C)(\epsilon_{\bar
     Q'}(p-\Delta)+m_{\bar Q'})}\right)\cr
&&+\frac12\frac{{\bf
      \Delta}^2-(E_B+M_B)(E_C-M_C)}{(E_C+M_C)(\epsilon_{\bar
     Q'}(p-\Delta)+m_{\bar Q'})(\epsilon_Q(q)+m_Q)}\Biggr\}\cr
&&+q_+\Biggl\{\frac1{\epsilon_Q(q)+m_Q}\Biggl(1+
  \frac12\frac{{\bf
      \Delta}^2-(E_B+M_B)(E_C-M_C)}{(E_B+M_B)}\cr
&&\times\left(\frac1{\epsilon_q(q)+m_q}+\frac1{\epsilon_{
      Q}(q-\Delta)+m_{Q})}\right)\Biggr)-\frac12\frac{{\bf
      \Delta}^2+(E_B-M_B)(E_C-M_C)}{(E_B+M_B)(\epsilon_{
      Q}(p)+m_{Q})}\cr
&&\times\left(\frac1{\epsilon_q(q)+m_q}-\frac1{\epsilon_Q(q-\Delta)+m_Q}\right)\Biggr\}\Biggr]
+\frac{{\bf
      \Delta}^2+(E_B-M_B)(E_C-M_C)}{E_B^2}\cr
&&\times\left(\frac{E_C+M_C}{E_B+M_B}
-\frac{E_B^2}{{\bf\Delta}^2+(E_B+M_B)(E_C+M_C)}\right){\cal M}_{\rm Coul}({\bf p},{\bf q}),
\end{eqnarray}
where the $p_+,q_+$ momenta are given by
$p_+=1/2(p_x+ip_y)=-\sqrt{\frac{2\pi}3}pY_{11}(\Omega)$ and
$q_+=1/2(q_x+iq_y)=-\sqrt{\frac{2\pi}3}qY_{11}(\Omega)$,
and the final meson energies are
$E_I=\sqrt{M_I^2+{\bf \Delta}^2}$ ($I=B,C$). 

The expressions for the contribution of the $(d2)$ diagram in
Fig.~\ref{fig:1a} can be obtained from
Eqs.~(\ref{eq:msc})-(\ref{eq:mtc}) by 
the obvious replacements
($Q\leftrightarrow Q'$, $M_B\leftrightarrow M_C$).  

\section{Results}
\label{sec:rd}

The differential decay rate of the strong decay $A\to BC$ is expressed
through the decay amplitude $h_{fi}$ by \cite{abs} 
\begin{equation}
  \label{eq:dg}
  \frac{d\Gamma_{A\to BC}}{d\Omega}=2\pi\frac{|{\bf
      \Delta}|E_BE_C}{M_A}|h_{fi}|^2,
\end{equation}
where the modulus of the recoil momentum of the final mesons is given by
$$|{\bf \Delta}|=\frac{\sqrt{[M_A^2-(M_B+M_C)^2][M_A^2-(M_B-M_C)^2]}}{2M_A}.$$

Now we substitute the relativistic meson wave functions, obtained
previously in the calculations of meson mass spectra 
\cite{mass}, into Eqs.~(\ref{eq:hh}),
(\ref{eq:isp1})--(\ref{eq:dg}) and determine the corresponding decay
amplitudes and decay rates. 
The results for the strong decay rates of vector ($^3S_1$)
mesons into a
pair of pseudoscalar ($^1S_0$) mesons  are given in
Table~\ref{tab:dr} in comparison with the predictions of the $^3P_0$
model  \cite{abs,bbp,sef}, 
the microscopic model \cite{sef2}, the Dyson-Schwinger (DS) equation
model \cite{ikr}, the Nambu-Jona-Lasinio (NJL) quark model \cite{venk}  and available
experimental data \cite{pdg}. The results are presented for the decays of
light ($\rho$, $\phi$, $K^*$), heavy-light ($D^*$, $D_s^*$) mesons and heavy
($\Psi$, $\Upsilon$) quarkonia. Note that in our
calculations we 
consistently take into account the
relativistic structure of the strong decay
amplitudes, transformations of the meson wave functions from the rest to
the
moving reference frame as well as the corrections to the rest frame
wave functions originating from the relativistic contributions to the
quark-antiquark interaction potential, which are treated
nonperturbatively in our model.  We find that our predictions agree
well with the available experimental data.   

\begin{table}
  \caption{Strong decay rates of vector ($^3S_1$)  mesons into a
pair of pseudoscalar ($^1S_0$) mesons  (in MeV).} 
  \label{tab:dr}
\begin{ruledtabular}
\begin{tabular}{cccccccc}
Decay& $\Gamma_{\rm our}$  & $\Gamma_{^3P_0}$  \cite{abs,bbp}&
$\Gamma_{^3P_0}$   \cite{sef,sef2}&
$\Gamma_{\rm mic.}$   \cite{sef2}&$\Gamma_{\rm DS}$   \cite{ikr}&$\Gamma_{\rm NJL}$   \cite{venk}& $\Gamma_{\rm exp}$ \cite{pdg}\\
\hline
$\rho\to \pi\pi$ & 124 &79 & 160 && 118&149&$147.8\pm0.9$\\
$\phi\to K\bar K$ & 3.3 & 2.5 &    & &&4.2&$3.55\pm 0.05$\\
$K^*\to K\pi$ & 46 & 21  &    & &52&51&$47.4\pm 0.6$\\
$D^*\to D\pi$ & 0.062 &0.025  & 0.036 & &0.038&0.063&$0.082\pm 0.002$\\
$\rho(2S)\to \pi\pi$ & 160 &74 &  &&&22&\\
$\rho(2S)\to K\bar K$ & 14 &35 &  &&\\
$\phi(2S)\to K\bar K$ & 17 & 89 &    & &&10&\\
$D^*(2S)\to D\pi$ & 20 & 1 &  & &\\
$D^*(2S)\to D_sK$ & 2.6 & 0.1 &  & &\\
$D^*(2S)\to D\eta$ & 2.2 & 0.4 &  & &\\
$D_s^*(2S)\to DK$ & 21 &17  &  & &\\
$D_s^*(2S)\to D_s\eta$ & 1.4 &2.6  &  & &\\
$\Psi(3S)\to D\bar D$& 11 & 0.1 & 4.61 & 10.17\\
$\Psi(3S)\to D_s\bar D_s$& 4.0 &7.8  & 2.08 & 1.14\\
$\Upsilon(4S)\to B\bar B$& 18 &  & 20.59 && &&$20.5\pm 2.5$\\
$\Upsilon(5S)\to B\bar B$& 4.3 &  &  & &&&$3.0\pm 1.7$\\  
$\Upsilon(5S)\to B_s\bar B_s$& 0.3 &  &  && &&$0.28\pm 0.28$\\
\end{tabular}
 \end{ruledtabular}
\end{table}

It is interesting to analyze the role of the relativistic
contributions to the considered strong decay rates. We use the
$\rho\to\pi\pi$ decay as an example since both initial and final mesons
contain only light quarks and thus relativistic effects are very
important. Taking the nonrelativistic limit of expressions
(\ref{eq:msc})--(\ref{eq:mtc}) and calculating  the strong decay
rate (\ref{eq:dg}), we get $\Gamma^{\rm
  NR}(\rho\to\pi\pi)=331$~MeV. Omitting  contributions coming from
the Lorentz boost of the  meson wave functions in the relativistic
expressions (\ref{eq:msc})--(\ref{eq:mtc}) we get the $\rho\to\pi\pi$
decay rate of 145~MeV, while complete relativistic calculation gives
$\Gamma(\rho\to\pi\pi)=124$~MeV. Therefore we conclude that
the account of the relativistic structure of the decay amplitude
reduces the nonrelativistic $\rho\to\pi\pi$ decay rate by 56\%, while
the relativistic transformations of the meson wave functions give
additional reduction of 6.4\%. We can also analyze contributions of
the different Lorentz structures of the interaction potential to the
decay rate. The contributions of  potentials considered separately are the
following. The scalar confining interaction:
$\Gamma^S(\rho\to\pi\pi)=54$~MeV; the vector confining interaction:
$\Gamma^V(\rho\to\pi\pi)=106$~MeV; the one-gluon exchange (OGE):
$\Gamma^{\rm OGE}(\rho\to\pi\pi)=4.4$~MeV.        

In principle, the similar analysis can be done for other considered
decays. Here we additionally present results for the $\Psi(3S)\to D\bar D$ decay,
since it involves both light ($u,d$) and heavy ($c$) quarks. In the
nonrelativistic limit we get $\Gamma^{\rm NR}(\Psi(3S)\to D\bar
D)=18.6$~MeV. Account of the relativistic structure of the decay
amplitude reduces it by 38\%, while the relativistic transformations
of the meson wave functions give additional reduction of 2.7\%,
leading to
the final value $\Gamma(\Psi(3S)\to D\bar D)=11$~MeV. The
contributions of the Lorentz structures of the interaction potential are
now the following: $\Gamma^S(\Psi(3S)\to D\bar D)=32$~MeV;
$\Gamma^V(\Psi(3S)\to D\bar D)=18$~MeV; $\Gamma^{\rm OGE}(\Psi(3S)\to
D\bar D)=0.23$~MeV. We see that, as naively expected, the role of the relativistic effects
is somewhat reduced for the strong decays of heavy mesons, especially
the ones coming from the recoil of final mesons (\ref{wig}).

The results of Refs.~\cite{abs,bbp}
are based on the  $^3P_0$ model with the universal flavour independent
strength parameter $\gamma$, while in Refs.~\cite{sef,sef2} the authors take into
account the scaling of $\gamma$ with the reduced mass of the
quark-antiquark pair in the decaying meson. The account for such scaling,
which was found to be logarithmic in the reduced mass, improves
agreement of the predictions with experimental data, 
but introduces an
additional free parameter, which determines the scaling. In all these
calculations 
Gaussian wave functions were used. 

The main advantage of
the microscopic approach consists in the fact that it completely
determines the strong decay dynamics without introducing a
strength parameter responsible for the production of the 
light quark pair.  In Ref.~\cite{abs}
the microscopic model was used for  the decay rate
of the $\rho$ meson with the result $\Gamma_{\rho\to\pi\pi}=243$~MeV,
which is too large relative to data. The possible sources of this
overestimate were attributed to the nonrelativistic consideration and the
choice of the $q\bar q$ wave functions in a
simple harmonic oscillator
form. Our calculations confirm this conjecture. As a result, the
prediction for the $\rho\to\pi\pi$ decay rate is reduced 
by almost a factor of two in fair agreement with data.  In Ref.~\cite{sef2}
the microscopic model was applied to
strong charmonium decays. The
employed model uses for the quark-antiquark interaction the sum of
one-gluon exchange, a nonperturbative confining term 
with scalar/vector Lorentz structure including
phenomenological string-breaking effects and 
(in case of light mesons)
Goldstone-boson exchange potentials. Calculations of the strong decays
were carried out in the nonrelativistic limit. We find that our
relativistic prediction for the $\Psi(3S)\to D\bar D$ decay rate agree
well with the result of Ref.~\cite{sef2}, while the one for the
$\Psi(3S)\to D_s\bar D_s$ decay rate differ by almost a factor of
four. The possible origin of this discrepancy can be attributed to
the difference in the wave functions and to the treatment of the
relativistic effects. As it was noted in Ref.~\cite{yopr}, this decay
should be very sensitive to the position of nodes of the $\Psi(3S)$
wave function and the resulting nodes in the decay amplitude.   

We find a reasonable agreement of our results with the predictions of
the relativistic quark model based on the DS equation \cite{ikr} and
the NJL model
\cite{venk}. The only exception is the significant difference of our
and the NJL model \cite{venk} values for the $\rho(2S)\to\pi\pi$
decay. Note that the $^3P_0$ model  \cite{bbp} gives the
intermediate result. Therefore experimental measurement of this decay
rate can help to discriminate theoretical approaches.

Strong decays of the $D^*$ mesons deserve a special attention, since
the phase space for their decays to the $D$ meson and $\pi$ is
small. Indeed, decays of the charged $D^*(2010)^+$ meson are
kinematically allowed both to $D^0\pi^+$ and $ D^+\pi^0$, while the neutral
$D^*(2007)^0$ meson can strongly decay only to $D^0\pi^0$. Due to the
strong phase space suppression of strong decays of the $D^*$ mesons,
their radiative decays start to play an
important role. Such radiative
decays were calculated in our paper \cite{radD} with the
comprehensive account of relativistic effects. In Table~\ref{tab:drD}
we confront our predictions for the decay rates  and branching
fractions of the charged and neutral $D^*$ mesons with previous
calculations based on the DS equation \cite{ikr} and the relativistic confinement quark
model (QCM) \cite{iv} and   available
experimental data. Good agreement of theoretical results and data is
observed. Note that in Ref.~\cite{ycclly} the strong decay of the
charged $D^{*+}$ meson to neutral $D^0$ meson and pion was considered in the quark model which
incorporates heavy quark symmetry and chiral dynamics. The prediction
for the decay rate $\Gamma(D^{*+}\to D^0\pi^+)=100$~keV (for the
favoured value of the constant $f$) was obtained which
is somewhat larger than the measured rate.

\begin{table}
  \caption{Decay rates  and branching fractions of the $D^*$ mesons.} 
  \label{tab:drD}
\begin{ruledtabular}
\begin{tabular}{ccccccc}
Decay& \multicolumn{2}{c}{$\Gamma$ (keV)}  & \multicolumn{4}{c}{$Br$ }\\
\cline{2-3} \cline{4-7}
& our & Experiment \cite{pdg}& our& DS \cite{ikr}&QCM \cite{iv}& Experiment \cite{pdg}\\
\hline
$D^*(2010)^+\to D^0\pi^+$ & 42 &$56\pm 1.5$ &0.667&0.683&0.687& $0.677\pm0.005$\\
$\phantom{D^*(2010)^+}\to D^+\pi^0$ & 20 &$25.6\pm 0.9$&0.317&0.316&0.309&
$0.307\pm0.005$\\
$\phantom{D^*(2010)^+}\to D^+\gamma$\ \ \ & 1.04 &$1.3\pm 0.5$&0.016&0.001&0.004&
$0.016\pm0.004$\\
Total & 63& $83.4\pm1.8$\\
\\
$D^*(2007)^0\to D^0\pi^0$ & 19 & &0.623& 0.826&0.682&$0.619\pm0.029$\\
$\phantom{D^*(2007)^0}\to D^0\gamma$\ \ \ & 11.5 &&0.377&0.174&0.318&
$0.381\pm0.029$\\
Total & 30.5& $<2100$\\
\end{tabular}
 \end{ruledtabular}
\end{table}

\section{Conclusions}
\label{sec:concl}

In this paper we propose the relativistic extension of the microscopic
model of strong meson decays. It is developed in the framework of
the relativistic quark model based on the quasipotential approach and
QCD-motivated interquark potential. This model was previously
successfully applied to the
calculation of hadron spectroscopy and radiative and weak decays. Such
approach allowed us to get the wave functions of light and heavy mesons
with the nonperturbative account of the relativistic
effects. The consistent relativistic approach for the calculation of
the decay matrix elements is now applied for the calculation of the strong
decay amplitudes. The relativistic transformation of meson wave
functions from rest to moving reference frames 
is explicitly taken into
account. The obtained decay matrix elements are
treated without application of the nonrelativistic expansion. Here we
test our approach in calculating strong decays of the vector ($^3S_1$)
mesons to the pair of pseudoscalar ($^1S_0$) mesons. The presence of
only $S$-wave mesons in the initial and final states significantly
simplifies the angular integration in the decay matrix elements.      
The decay rates are obtained for decays of light, heavy-light 
mesons and heavy quarkonia. All calculations are performed with all
model parameters kept
fixed from previous calculations of meson spectroscopy. No additional
parameters are necessary for describing  the production of the
light $q\bar q$ pair, since it is considered to originate from the
same interaction term
as the interquark potential. The obtained results are
confronted with calculations 
within the $^3P_0$ model, the DS equation model, the NJL model  and nonrelativistic microscopic models as well as
available experimental data. The overall agreement of the obtained results
with experiment is found.  

\acknowledgments
The authors are grateful to 
M.~A.~Ivanov,  V. A. Matveev, D. I. Melikhov and V. I. Savrin  
for  useful discussions.
This work was supported in part by the {\it Russian
Foundation for Basic Research} under Grant No.12-02-00053-a.

\end{document}